\documentclass[11pt]{article}
\usepackage[body={16.6cm,23cm}]{geometry}
\usepackage[hang,small,center]{caption}
\usepackage{amsmath,amssymb}
\usepackage{amsfonts}
\usepackage{mathrsfs}
\usepackage{dsfont}

\usepackage{epic}
\usepackage{eepic}
\usepackage{graphicx}

\usepackage{pgf,pgfarrows}

\usepackage{cite}
\usepackage{enumerate}
\usepackage[curve]{xy}
\newcommand{\ds}{\displaystyle}
\renewcommand{\author}[1]{\large\rm #1\\ \bigskip}
\newcommand{\address}[1]{{\normalsize\it #1\\}\bigskip}
\renewcommand{\title}[1]{\bigskip\bigskip\Large\bf #1\bigskip\bigskip\\}

\newcommand{\Bigpsi}[3]{\phantom{\Psi}_2 \kern -.05em
\Psi_2\left(\genfrac{}{}{0pt}{}{#1}{#2}\biggl|#3\right)}

\newcommand{\bea}{\begin{eqnarray}}
\newcommand{\eea}{\end{eqnarray}}

\newcommand{\beq}{\begin{equation}}
\newcommand{\eeq}{\end{equation}}

\newcommand{\ii}{\mathsf{i}}

\newcommand{\qq}{{\mathsf q}}
\newcommand{\pp}{{\mathsf p}}

\newcommand{\iW}{\mathbb{W}}
\newcommand{\iS}{\mathbb{S}}

\newcommand{\iV}{\mathbb{V}}

\def\EXP{\textrm{{\large e}}}

\newcommand{\url}[1]{}

\renewcommand{\textcolor}[1]{}

\newcounter{app}
\newcounter{sapp}[app]

\begin{document}

\vglue 2 cm
\begin{center}
\title{Comment on star-star relations in statistical mechanics and elliptic
  gamma-function identities}  
\author{Vladimir V.~Bazhanov$^{1,2}$, \ Andrew P.~Kels$^{1}$ and \ Sergey M.~Sergeev$^3$}
\address{$^1$Department of Theoretical Physics,
         Research School of Physics and Engineering,\\
    Australian National University, Canberra, ACT 0200, Australia.\\\ \\
$^2$Mathematical Sciences Institute,\\
      Australian National University, Canberra, ACT 0200,
      Australia.\\\ \\
$^3$Faculty of Education Science Technology \& Mathematics,\\
University of Canberra, Bruce ACT 2601, Australia.}

\vspace{2cm}

\begin{abstract} 
We prove a recently conjectured star-star relation, which plays the
role of an integrability condition for a class of 2D Ising-type models
with multicomponent continuous spin variables. Namely, we reduce this
relation to an identity for elliptic gamma
functions, previously obtained by Rains.  
\end{abstract}

\end{center}

\newpage
\section{Introduction}
Recently two of us \cite{Bazhanov:2010kz, Bazhanov:2011mz} introduced a
new class of exactly solvable 2D lattice models of statistical
mechanics, which involve continuous spin variables taking values on a
circle. The interest in these models is motivated by various 
applications. In statistical mechanics they
serve as rather general ``master models'', containing  many important
particular 
limits, such as the Ising, chiral Potts \cite{BPY87,AuY87},
Kashiwara-Miwa  \cite{Kashiwara:1986}, Faddeev-Volkov
\cite{FV95,BMS07a} and other models.
Mathematically, the new models 
are deeply related to the theory of elliptic hypergeometric functions
\cite{Spiridonov-essays}.  
For instance, 
the celebrated
elliptic beta integral \cite{Spiridonov-beta}, 
which lies at the basis of this theory,
is shown \cite{Bazhanov:2010kz} 
to be a Yang-Baxter (star-triangle) relation, defining 
perfectly physical integrable lattice models of statistical mechanics.
Other interesting connections are discussed 
in \cite{Spiridonov:2010em,Kashaev:2012cz,Chicherin:2012yn,Chicherin:2012jz}. 
We mention, in particular, that the elliptic gamma-functions arise in
calculations of  
superconformal indices connected with 
electric-magnetic dualities in 4D ${\cal N}=1$ superconformal
Yang-Mills theories \cite{Spiridonov:2011hf}. Most remarkably,
as recently discovered in \cite{Yamazaki:2012cp,  Terashima:2012cx,
  Xie:2012mr}, the superconformal indices in 
4D superconformal quiver gauge theories precisely coincide with
partition functions of the 2D lattice ``master models'' 
\cite{Bazhanov:2010kz, 
  Bazhanov:2011mz} discussed here. Interestingly, in this
correspondence the Seiberg  duality for the superconformal indices
reduces to Baxter's $Z$-invariance \cite{Bax1} for the partition function
under (generalized) ``star-triangular moves'' of the 2D lattice. 

Here we resolve an outstanding question for these models 
concerning the so-called {\em
  star-star relation},  conjectured in \cite{Bazhanov:2011mz}. 
This relation serves as an integrability condition,
as it implies the Yang-Baxter equation, the $Z$-invariance of the
partition function and the commutativity of row-to-row
transfer matrices. In this letter we completely prove this star-star  
relation by reducing it to a transformation formula for elliptic
hypergeometric integrals, previously obtained by Rains
\cite{Rains:2010}. 

\section{Edge-interaction model with continuous spins}
\begin{figure}[h]
\vspace{1.cm}
\centering
\setlength{\unitlength}{1cm}
\begin{picture}(8,6)
\put(1,1){\begin{picture}(7,5)
 \thinlines
 \dottedline{0.08}(1,0)(1,5)\path(0.9,4.9)(1,5)(1.1,4.9)\put(0.9,-0.3){\scriptsize $v'$}
 \dottedline{0.08}(3,0)(3,5)\path(2.9,4.9)(3,5)(3.1,4.9)\put(2.9,-0.3){\scriptsize $v'$}
 \dottedline{0.08}(5,0)(5,5)\path(4.9,4.9)(5,5)(5.1,4.9)\put(4.9,-0.3){\scriptsize $v'$}
 \dottedline{0.08}(0,2)(7,2)\path(6.9,1.9)(7,2)(6.9,2.1)\put(-0.3,1.9){\scriptsize $u'$}
 \dottedline{0.08}(0,4)(7,4)\path(6.9,3.9)(7,4)(6.9,4.1)\put(-0.3,3.9){\scriptsize $u'$}
 \drawline(2,0)(2,5)\path(1.9,4.9)(2,5)(2.1,4.9)\put(1.9,-0.3){\scriptsize $v$}
 \drawline(4,0)(4,5)\path(3.9,4.9)(4,5)(4.1,4.9)\put(3.9,-0.3){\scriptsize $v$}
 \drawline(6,0)(6,5)\path(5.9,4.9)(6,5)(6.1,4.9)\put(5.9,-0.3){\scriptsize $v$}
 \drawline(0,1)(7,1)\path(6.9,0.9)(7,1)(6.9,1.1)\put(-0.3,0.9){\scriptsize $u$}
 \drawline(0,3)(7,3)\path(6.9,2.9)(7,3)(6.9,3.1)\put(-0.3,2.9){\scriptsize $u$}
 \Thicklines
 \path(1.5,0.5)(0.55,1.45)
 \path(0.55,1.55)(1.5,2.5)
 \path(1.5,2.5)(0.55,3.45)
 \path(0.55,3.55)(1.5,4.5)
 \path(1.5,0.5)(2.45,1.45)
 \path(2.45,1.55)(1.5,2.5)
 \path(1.5,2.5)(2.45,3.45)
 \path(2.45,3.55)(1.5,4.5)
 \path(3.5,0.5)(2.55,1.45)
 \path(2.55,1.55)(3.5,2.5)
 \path(3.5,2.5)(2.55,3.45)
 \path(2.55,3.55)(3.5,4.5)
 \path(3.5,0.5)(4.45,1.45)
 \path(4.45,1.55)(3.5,2.5)
 \path(3.5,2.5)(4.45,3.45)
 \path(4.45,3.55)(3.5,4.5)
 \path(5.5,0.5)(4.55,1.45)
 \path(4.55,1.55)(5.5,2.5)
 \path(5.5,2.5)(4.55,3.45)
 \path(4.55,3.55)(5.5,4.5)
 \path(5.5,0.5)(6.45,1.45)
 \path(6.45,1.55)(5.5,2.5)
 \path(5.5,2.5)(6.45,3.45)
 \path(6.45,3.55)(5.5,4.5)
 \put(1.5,0.5){\circle*{0.15}}
 \put(3.5,0.5){\circle*{0.15}}
 \put(5.5,0.5){\circle*{0.15}}
 \put(0.5,1.5){\circle{0.15}}
 \put(2.5,1.5){\circle{0.15}}
 \put(4.5,1.5){\circle{0.15}}
 \put(6.5,1.5){\circle{0.15}}
 \put(1.5,2.5){\circle*{0.15}}
 \put(3.5,2.5){\circle*{0.15}}
 \put(5.5,2.5){\circle*{0.15}}
 \put(0.5,3.5){\circle{0.15}}
 \put(2.5,3.5){\circle{0.15}}
 \put(4.5,3.5){\circle{0.15}}
 \put(6.5,3.5){\circle{0.15}}
 \put(1.5,4.5){\circle*{0.15}}
 \put(3.5,4.5){\circle*{0.15}}
 \put(5.5,4.5){\circle*{0.15}}
 \end{picture}}
 \end{picture}
 \caption{The square lattice shown with bold sites and bold edges drawn
   diagonally. The associated medial lattice is drawn with thin and
   dotted horizontal and vertical lines. The lines are oriented and
   carry rapidity variables $u$, $u'$, $v$ and $v'$.}
\label{fig-lattice}
\end{figure}
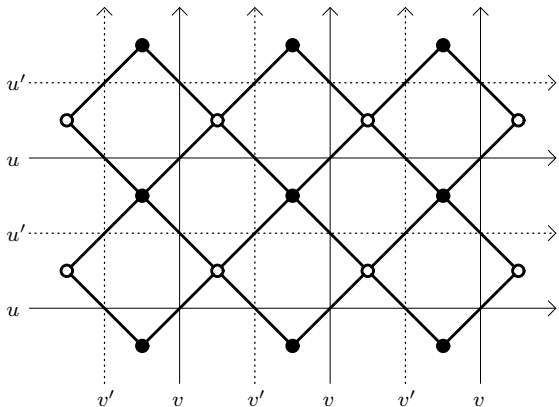

In this Section we formulate the star-star relation conjectured in
\cite{Bazhanov:2011mz} (see \eqref{star-star} below).
First, we need to briefly describe the associated two-dimensional
edge-interaction model; full details can be found in 
\cite{Bazhanov:2011mz}.  
Consider the regular square lattice, drawn diagonally as in
Fig.~\ref{fig-lattice}.
The edges of the lattice are shown with bold lines
and the sites are shown with either open or filled circles in a
checkerboard order. At the moment we will not distinguish these two type
of sites.
At each lattice  site place a $n$-component continuous spin variable
\begin{equation}
\boldsymbol{x}=\{x_1,\dots,x_n\}\in\mathbb{R}^n, \qquad 
0\le x_j< \pi,\qquad \sum_{j=1}^n x_j = 0\pmod \pi\;. \label{spinvar}
\end{equation}
Note that due to the restriction on the total sum, there are only
$(n-1)$ independent variables $x_j$. For further reference define the
integration measure
\begin{equation}\label{measure}
\int d\boldsymbol{x}=\int_0^{\pi}\cdots \int_0^{\pi} dx_1\cdots
dx_{n-1}\;, 
\end{equation}
over the space of states of a single spin.
Fig.~\ref{fig-lattice} also shows an auxiliary {\em medial}\/ lattice
whose sites lie on the edges of the original square lattice.
The medial lattice is drawn with alternating
thin and dotted lines. The lines are directed as indicated by  arrows.
To each horizontal (vertical) line on the medial lattice assign
a rapidity variable $u$ ($v$). In general these variables may be
different for different lines. 
However, a convenient level
of generality that we shall use here is
to assign the same rapidity $u$ to all thin horizontal lines
and the same variable $u'$ to all dotted horizontal lines. Similarly,
assign the variables $v$ and $v'$ to thin and dotted
vertical lines as indicated in
Fig.~\ref{fig-lattice}.

Two spins interact only if they are connected by an edge.
To define the Boltzmann weights we need to introduce 
the elliptic gamma-function
\cite{Spiridonov-essays}.
Let $\qq,\pp$ be two elliptic nomes (they play the role of
the temperature-like parameters),
\begin{equation}
\pp\;=\;\EXP^{\ii\pi\sigma}\;,\quad
\qq=\EXP^{\ii\pi\tau}\;,\quad \mbox{Im}\,\sigma>0,  \quad \mbox{Im}\,\tau>0\,, 
\end{equation}
and 
\beq
\eta=-\ii\pi(\sigma+\tau)/2\,,
\eeq
denote the ``crossing parameter''.
Define the elliptic gamma-function\footnote{Our function $\Phi(z)$
  coincides with $\Gamma(\EXP^{-2\ii (z-\eta)}; \pp^2,\qq^2)$ in the
  notation of ref. \cite{Spiridonov-essays}.}
\begin{equation}
\Phi(z)\;=\;\prod_{j,k=0}^\infty\frac{1-\EXP^{2\ii
z}\qq^{2j+1}\pp^{2k+1}}{1-\EXP^{-2\ii
z}\qq^{2j+1}\pp^{2k+1}}\;=\;\exp\left\{\sum_{k\neq 0}
\frac{\EXP^{-2\ii
zk}}{k(\qq^k-\qq^{-k})(\pp^k-\pp^{-k})}\right\}\;,\label{Phi-def}
\end{equation}
where the product formula is valid for all $z$, while the
exponential formula is only valid in the strip
\begin{equation}
-\textrm{Re}\,\eta < \textrm{Im}\,z < \textrm{Re}\,\eta\;.
\end{equation}
The function \eqref{Phi-def} possesses simple periodicity and
``reflection'' properties
\beq
\Phi(z+\pi)=\Phi(z)\,,
\qquad \Phi(z)\,\Phi(-z)=1\,.\label{reflection}
\eeq

Each edge is assigned a Boltzmann weight which depends
on spins at the ends of the edge and on two rapidities passing
through the edge.  
\begin{figure}[h]
\begin{center}
\setlength{\unitlength}{1cm}
\begin{picture}(15,4)
\put(0.5,1.5)
{\begin{picture}(6,2)
 \thinlines
 \drawline(0,0)(2,2)\path(2,1.9)(2,2)(1.9,2)
 \put(0,-0.3){\scriptsize $u$}
 \dottedline{0.08}(2,0)(0,2)\path(0,1.9)(0,2)(0.1,2)
 \put(2,-0.3){\scriptsize $v'$}
 \Thicklines
 \path(0,1)(2,1)
 \put(0,1){\circle*{0.1}}
 \put(2,1){\circle*{0.1}}
 \put(-0.4,0.9){\scriptsize $x$}
 \put(2.2,0.9){\scriptsize $y$}
 \put(0.3,-1.0){\scriptsize $\iW_{u-v'}(\boldsymbol{x},\boldsymbol{y})$}
 \thinlines
 \drawline(4,0)(6,2)\path(6,1.9)(6,2)(5.9,2)
 \drawline(6,0)(4,2)\path(4,1.9)(4,2)(4.1,2)
 \put(4,-0.3){\scriptsize $u$}
 \put(6,-0.3){\scriptsize $v$}
 \Thicklines
 \path(5,0)(5,2)
 \put(5,0){\circle*{0.1}}
 \put(5,2){\circle*{0.1}}
 \put(4.9,-0.3){\scriptsize $y$}
 \put(4.9,2.2){\scriptsize $x$}
 \put(4.3,-1.0){\scriptsize $\overline{\iW}_{u-v}(\boldsymbol{y},\boldsymbol{x})$}
\end{picture}}
\put(8,1.5){\begin{picture}(6,2)
 \thinlines
 \dottedline{0.08}(0,0)(2,2)\path(2,1.9)(2,2)(1.9,2)
 \put(0,-0.3){\scriptsize $u'$}
 \drawline(2,0)(0,2)\path(0,1.9)(0,2)(0.1,2)
 \put(2,-0.3){\scriptsize $v$}
 \Thicklines
 \path(0,1)(2,1)
 \put(0,1){\circle*{0.1}}
 \put(2,1){\circle*{0.1}}
 \put(-0.4,0.9){\scriptsize $y$}
 \put(2.2,0.9){\scriptsize $x$}
 \put(0.3,-1.0){\scriptsize $\iW_{u'-v}(\boldsymbol{x},\boldsymbol{y})$}
 \thinlines
 \dottedline{0.08}(4,0)(6,2)\path(6,1.9)(6,2)(5.9,2)
 \dottedline{0.08}(6,0)(4,2)\path(4,1.9)(4,2)(4.1,2)
 \put(4,-0.3){\scriptsize $u'$}
 \put(6,-0.3){\scriptsize $v'$}
 \Thicklines
 \path(5,0)(5,2)
 \put(5,0){\circle*{0.1}}
 \put(5,2){\circle*{0.1}}
 \put(4.9,-0.3){\scriptsize $x$}
 \put(4.9,2.2){\scriptsize $y$}
 \put(4.3,-1.0){\scriptsize $\overline{\iW}_{u'-v'}(\boldsymbol{y},\boldsymbol{x})$}
\end{picture}}
\end{picture}
\caption{Four different types of edges and their Boltzmann weights.}
\label{fig-crosses}
\end{center}
\end{figure}
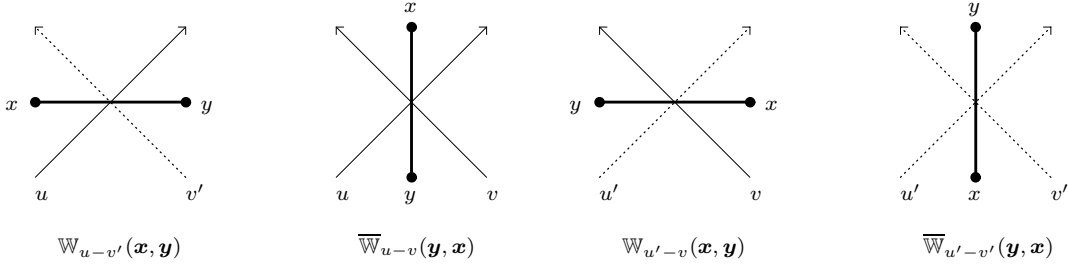
There are four types of edges differing by
orientations and types of the directed rapidity lines passing through
the edge. They are assigned different Boltzmann weights
as shown in Fig.\ref{fig-crosses}. These weights are defined as
\begin{equation}
\iW_\alpha(\boldsymbol{x},\boldsymbol{y})
=\kappa_n(\alpha)^{-1}\prod_{j,k=1}^{n}\Phi(x_j-y_k+\ii\alpha),\quad  
\overline{\iW}_\alpha(\boldsymbol{x},\boldsymbol{y})
=\sqrt{\iS(\boldsymbol{x})\iS(\boldsymbol{y})}\iW_{\eta-\alpha}(x,y)\;,
\label{weights}
\end{equation}
where $\kappa_n(\alpha)$ is a normalization factor. 
The single-spin function $\iS$ is given by
\begin{equation}
\iS(\boldsymbol{x})\;=\;\varkappa_s^{-1}
\ds\prod_{j\neq k}
  \Big\{\Phi(x_j-x_k+\ii\eta)\Big\}^{-1}\,,\qquad 
\varkappa_s\;=\;{n!}\left(\frac{\pi}
{G(\qq)\,G(\pp)}\right)^{n-1}\,,
 \label{S-def}
\end{equation}
where the indices $j,k$ run over the values $1,2,\ldots,n$ and
\beq
G(z)=\prod_{k=1}^\infty(1-z^{2k})\,.\label{G-def}
\eeq 
The partition function is defined as
\beq
Z=\int  \ \prod_{\langle\boldsymbol{x}\boldsymbol{y}\rangle}  
\iW_{u-v'}(\boldsymbol{x},\boldsymbol{y})\ 
\prod_{\langle\boldsymbol{y}\boldsymbol{x}\rangle}
\overline{\iW}_{u-v}(\boldsymbol{y},\boldsymbol{x})\ 
\prod_{\langle\boldsymbol{x}\boldsymbol{y}\rangle}  
\iW_{u'-v}(\boldsymbol{x}\boldsymbol{y})\ 
\prod_{\langle\boldsymbol{y},\boldsymbol{x}\rangle}
\overline{\iW}_{u'-v'}(\boldsymbol{y},\boldsymbol{x})\ 
\prod_{\rm sites} d \boldsymbol{x}, \label{Z-edge}
\eeq
where the four products are taken, respectively, over the four types
of edges shown in 
Fig.~\ref{fig-crosses}. The integral is 
taken over all configurations of the spin variables on the internal
lattice sites. The boundary spins are kept fixed.

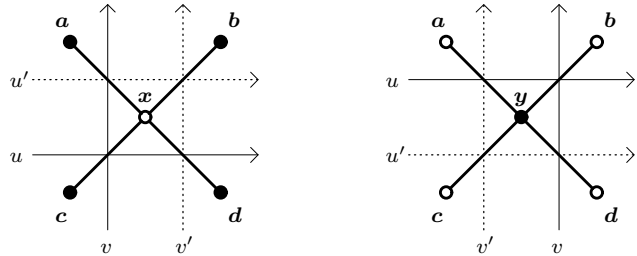
\begin{figure}[ht]
\begin{center}
\setlength{\unitlength}{1cm}
\begin{picture}(10,4)
\put(1,1){\begin{picture}(3,3)
 \thinlines
% \color{red}
 \drawline(0.5,-0.5)(0.5,2.5)\path(0.4,2.4)(0.5,2.5)(0.6,2.4)
 \drawline(-0.5,0.5)(2.5,0.5)\path(2.4,0.4)(2.5,0.5)(2.4,0.6)
 \put(0.4,-0.8){\scriptsize $v$}
 \put(-0.8,0.4){\scriptsize $u$}
% \color{blue}
 \dottedline{0.08}(1.5,-0.5)(1.5,2.5)\path(1.4,2.4)(1.5,2.5)(1.6,2.4)
 \dottedline{0.08}(-0.5,1.5)(2.5,1.5)\path(2.4,1.4)(2.5,1.5)(2.4,1.6)
 \put(1.4,-0.8){\scriptsize $v'$}
 \put(-0.8,1.4){\scriptsize $u'$}
% \color{black}
 \Thicklines
\path(0,0)(.95,.95)
 \path(1.05,1.05)(2,2)
 \path(1.05,.95)(2,0)
 \path(.95,1.05)(0,2)
 \put(1,1){\circle {0.15}}
 \put(0,0){\circle*{0.15}}
 \put(0,2){\circle*{0.15}}
 \put(2,0){\circle*{0.15}}
 \put(2,2){\circle*{0.15}}
 \put(0.9,1.2){\scriptsize $\boldsymbol{x}$}
 \put(-0.2,2.2){\scriptsize $\boldsymbol{a}$}
 \put(-0.2,-0.4){\scriptsize $\boldsymbol{c}$}
 \put(2.1,2.2){\scriptsize $\boldsymbol{b}$}
 \put(2.1,-0.4){\scriptsize $\boldsymbol{d}$}
\end{picture}}
\put(6,1){\begin{picture}(3,3)
 \thinlines
% \color{blue}
 \dottedline{0.08}(0.5,-0.5)(0.5,2.5)\path(0.4,2.4)(0.5,2.5)(0.6,2.4)
 \dottedline{0.08}(-0.5,0.5)(2.5,0.5)\path(2.4,0.4)(2.5,0.5)(2.4,0.6)
 \put(0.4,-0.8){\scriptsize $v'$}
 \put(-0.8,0.4){\scriptsize $u'$}
% \color{red}
 \drawline(1.5,-0.5)(1.5,2.5)\path(1.4,2.4)(1.5,2.5)(1.6,2.4)
 \drawline(-0.5,1.5)(2.5,1.5)\path(2.4,1.4)(2.5,1.5)(2.4,1.6)
 \put(1.4,-0.8){\scriptsize $v$}
 \put(-0.8,1.4){\scriptsize $u$}
% \color{black}
 \Thicklines
 \path(0.05,0.05)(.95,.95)
 \path(1.05,1.05)(1.95,1.95)
 \path(1.05,.95)(1.95,0.05)
 \path(.95,1.05)(0.05,1.95)
 \put(1,1){\circle*{0.15}}
 \put(0,0){\circle{0.15}}
 \put(0,2){\circle{0.15}}
 \put(2,0){\circle{0.15}}
 \put(2,2){\circle{0.15}}
 \put(0.9,1.2){\scriptsize $\boldsymbol{y}$}
 \put(-0.2,2.2){\scriptsize $\boldsymbol{a}$}
 \put(-0.2,-0.4){\scriptsize $\boldsymbol{c}$}
 \put(2.1,2.2){\scriptsize $\boldsymbol{b}$}
 \put(2.1,-0.4){\scriptsize $\boldsymbol{d}$}
\end{picture}}
\end{picture}
\caption{Two types of four-edge stars: a white-centred star
  $\mathbb{V}^{(1)}$ (left) and a black-centred star 
  $\mathbb{V}^{(2)}$ (right).}\label{fig-IRF}
\end{center}
\end{figure}

Note that the
lattice in Fig.~\ref{fig-lattice} can be formed by periodic
translations of a {\em four-edge star}, consisting of
four edges meeting at the same site.
A little  inspection shows that
there are only two different types of
such stars shown in Fig.~\ref{fig-IRF}.
They are either 
centred around ``white'' sites, shown with open
circles or around ``black'' sites, shown with filled circles.
Applying the rules of Fig.~\ref{fig-crosses} one can write
Boltzmann weights corresponding to these stars
\begin{equation}\label{V1}
\ds \mathbb{V}_{\boldsymbol{u}\boldsymbol{v}}^{(1)}\left(\begin{array}{cc}
\boldsymbol{a} & \boldsymbol{b}\\ \boldsymbol{c} & \boldsymbol{d}\end{array}\right)=
\int d\boldsymbol{x}\,
\overline{\iW}_{u-v}(\boldsymbol{c},\boldsymbol{x})\,
\overline{\iW}_{u'-v'}(\boldsymbol{b},\boldsymbol{x})\,
\iW_{u'-v}(\boldsymbol{x},\boldsymbol{a})\,
\iW_{u-v'}(\boldsymbol{x},\boldsymbol{d})\;,
\end{equation}
and
\begin{equation}\label{V2}
\ds \mathbb{V}_{\boldsymbol{u}\boldsymbol{v}}^{(2)}\left(\begin{array}{cc}
\boldsymbol{a} & \boldsymbol{b}\\ \boldsymbol{c} &
\boldsymbol{d}\end{array}\right)=\int d\boldsymbol{y}\, 
\overline{\iW}_{u-v}(\boldsymbol{y},\boldsymbol{b})\,  
\overline{\iW}_{u'-v'}(\boldsymbol{y},\boldsymbol{c})\,
\iW_{u'-v}(\boldsymbol{d},\boldsymbol{y})\,
\iW_{u-v'}(\boldsymbol{a},\boldsymbol{y})\;,
\end{equation}
where the bold symbols
$\boldsymbol{u}=[u,u']$ and $\boldsymbol{v}=[v,v']$ 
stand for the rapidity pairs. 
It turns out that the above two expressions are simply connected to
each other 
\begin{equation}
{\iW_{v'-v}(\boldsymbol{d},\boldsymbol{c})\,
\iW_{u'-u}(\boldsymbol{d},\boldsymbol{b})} 
\,\,\mathbb{V}_{\boldsymbol{u}\boldsymbol{v}}^{(1)}\left(\begin{array}{cc}
\boldsymbol{a} & \boldsymbol{b} \\
\boldsymbol{c} & \boldsymbol{d}\end{array}\right)\;=\;
{\iW_{v'-v}(\boldsymbol{b},\boldsymbol{a})\,
\iW_{u'-u}(\boldsymbol{c},\boldsymbol{a})}
\,\,
\mathbb{V}_{\boldsymbol{u}\boldsymbol{v}}^{(2)}\left(\begin{array}{cc}
\boldsymbol{a} & \boldsymbol{b} \\
\boldsymbol{c} & \boldsymbol{d}\end{array}\right)\;.\label{star-star}
\end{equation}
This is precisely the {\em star-star relation} conjectured
in \cite{Bazhanov:2011mz} (in the same  
paper it was also verified in a few orders of perturbation
theory in the parameters $\pp$ and $\qq$).  Below we will give a
complete proof of \eqref{star-star} by reducing it to a mathematical
identity, previously obtained by Rains \cite{Rains:2010}.
Apparently, the above star-star relation is the simplest condition for
the Boltzmann weights 
which ensures the integrability of the considered model\footnote{%
\label{foot3}%
For $n = 2$ the star-star relation \eqref{star-star} is just a
consequence of the star-triangle relation, Eq.(1.5) of
\cite{Bazhanov:2010kz}, which is equivalent to the elliptic beta
integral \cite{Spiridonov:2010em,Spiridonov-beta}.  However, for $n\ge3$ the
corresponding star-triangle relation apparently does not exist (at
least it is not known to the authors) and the star-star relation
\eqref{star-star} seems to be the simplest relation of this type.}.
In particular, it implies the commutativity of the row-to-row transfer
matrices. 

\section{Proof of the star-star relation}
Here we will use the standard notation for the
elliptic gamma-function, which is simply 
related to our definition \eqref{Phi-def},  
\beq
\Gamma(z;\pp^2,\qq^2)=\prod_{j,k=0}^\infty\ 
\frac{(1-\pp^{2 j+2}\,\qq^{2 k+2}/z)}{(1-\pp^{2 j}\,\qq^{2 k}\,z)}\,,\qquad
\label{gamrel}
\Phi(x)=\Gamma(\pp\qq\,\EXP^{-2\ii x};\pp^2,\qq^2)\,,\qquad
\pp\qq=\EXP^{-2 \eta}\,.
\eeq
In the following we will omit the nome arguments $\pp^2$ and $\qq^2$,
assuming that $\Gamma(z)\equiv \Gamma(z;\pp^2,\qq^2)$. 
Following Rains \cite{Rains:2010} introduce the following elliptic
hypergeometric 
integral\footnote{%
We follow Sect.4 of \cite{Rains:2010}, where we set
$Z\equiv1$. Moreover, our indices 
numerating parameters in \eqref{tu-var} start from $1$ instead of
$0$ in \cite{Rains:2010}.} 
\beq
{\bf I}_{A_{n-1}}^{(n-1)}(\{t_i\},\{s_i\}) \;\stackrel{\textrm{def}}{=}\;
\frac{1}{n!}\left(\frac{G(\pp)\, G(\qq)}{2\pi\ii}\right)^{n-1}
\int_{|z_k|=1} 
\frac{\prod_{k=1}^n\prod_{j=1}^{2n} \Gamma(t_j z_k)\,\Gamma(s_j/z_k)}{
  \prod_{k\not=\ell } \Gamma(z_k/z_\ell)}\,
\prod_{k=1}^{n-1} \frac{dz_k}{z_k}\,,\label{int1}
\eeq
which involve $4n$ independent parameters  
\beq
\{t_i\}=\{t_1,t_2,\ldots,t_{2n}\},\quad
\{s_i\}=\{s_1,s_2,\ldots,s_{2n}\},\quad |t_i|,|s_i|<1,\quad i=1,\ldots,2n\,,
\label{tu-var}
\eeq
where $n\ge2$ and the function $G(z)$ is defined in \eqref{G-def}.
 The indices $k$ and $\ell$ in the denominator of
\eqref{int1} run over the values $1,\dots,n$.
All integrations are taken over the unit circles $|z_k|=1$, and the
variable $z_n$ is determined by the constraint
\beq
z_1z_2\cdots z_n=1.\label{z-con}
\eeq
The theorem 4.1 of \cite{Rains:2010} (where we set $m=n$ and $Z=1$)
states the following transformation formula,
\beq
{\bf I}^{(n-1)}_{A_{n-1}}(\{t_i\},\{s_i\})=
\Big(\prod_{j,k=1}^{2n}\,\Gamma(t_j s_k) \Big)\ 
{\bf
  I}^{(n-1)}_{A_{n-1}}(\{\widetilde{t}_i\},\{\widetilde{u}_i\})
\label{rains-th} 
\eeq
where the new parameters $\{\widetilde{t}_i\}$ and
$\{\widetilde{s}_i\}$ in the RHS are given by 
\beq
\widetilde{t}_i=T^\frac{1}{n}\,{t_i}^{-1},\qquad\widetilde{u}_i=U^\frac{1}{n}\,
s_i^{-1}, \qquad T=\prod_{j=1}^{2n} \,t_j,\qquad U=\prod_{j=1}^{2n}\,
s_j\,,\quad i=1,2,\ldots,2n\,.\label{tu-tilda} 
\eeq

Consider now the star weight \eqref{V1} and make a change
of variables $z_j=\EXP^{+2\ii x_j}$ and 
 \beq
 t_j=\EXP^{-2(u-v)-2\ii c_j},\quad t_{n+j}
 =\EXP^{-2(u'-v')-2\ii b_j},\quad
 s_j=\EXP^{2(u'-v-\eta)+2\ii a_j}\;,\quad
 s_{n+j}=\EXP^{2(u-v'-\eta)+2\ii d_j}\;,\label{t-change}
 \eeq
where $a_j,b_j,c_j,d_j$ are the components of
the spin variables and $j=1,\dots,n$. Note that due to \eqref{spinvar}
the new variables $z_j$ obey the constraint \eqref{z-con}.
Now taking into account that \eqref{S-def} can be written as 
\beq
\label{Schange}
\iS(\boldsymbol{x})=\kappa_s^{-1} \Big(\prod_{j\not= k}
\Gamma(z_j/z_k)\Big)^{-1},\qquad z_j=\EXP^{2\ii x_j}, 
\eeq
it is not difficult to check that
\beq
\mathbb{V}^{(1)}_{\boldsymbol{u}\boldsymbol{v}}\left(\begin{array}{cc}
  \boldsymbol{a} & \boldsymbol{b}\\ \boldsymbol{c} &
  \boldsymbol{d}\end{array}\right)\;=\;\varrho\; {\bf
  I}_{A_{n-1}}^{(n-1)}(\{t_i\},\{s_i\})\;, \label{V1I}
\eeq
where
\beq
\varrho=\frac{\sqrt{\mathbb{S}(\boldsymbol{c})\mathbb{S}(\boldsymbol{b})}}
{\kappa_n(\eta-u+v)\kappa_n(\eta-u'+v')\kappa_n(u'-v)\kappa_n(u-v')}\;,  
\label{varrho-def}  
\eeq
Further, using \eqref{gamrel} and the reflection property
\eqref{reflection} 
for the
elliptic gamma-function,   
one can re-write the ratio of the $\iW$-factors entering
\eqref{star-star} in the form 
\beq
{\iW_{v'-v}(\boldsymbol{b},\boldsymbol{a})\,
\iW_{u'-u}(\boldsymbol{c},\boldsymbol{a})}
\Big[{\iW_{v'-v}(\boldsymbol{d},\boldsymbol{c})\,
\iW_{u'-u}(\boldsymbol{d},\boldsymbol{b})}\Big]^{-1}=
\prod_{j,k=1}^{2n}\,\Gamma(t_j s_k) \label{W-factors}
\eeq
Next, substituting \eqref{t-change} into \eqref{tu-tilda} one gets
\beq
\widetilde{t}_j=\EXP^{-2(u'-v')+2\ii c_j}\;,\quad
\widetilde{t}_{n+j}=\EXP^{-2(u-v)+2\ii b_j}\;,\quad 
\widetilde{s}_j=\EXP^{2(u-v'-\eta)-2\ii a_j}\;,\quad
\widetilde{s}_{n+j}=\EXP^{2(u'-v-\eta)-2\ii d_j}\,. \label{tt-def}
\eeq
Consider now the star weight \eqref{V2} and make a change
of variables $z_j=\EXP^{-2\ii x_j}$ (note the minus sign in the
exponent). Using the variables \eqref{tt-def} one obtains
\begin{align}
\iV^{(2)}_{\boldsymbol{uv}}\left(\begin{array}{cc} \boldsymbol{a} &
  \boldsymbol{b}\\ \boldsymbol{c} &
  \boldsymbol{d}\end{array}\right)
=\varrho\  {\bf I}^{(n-1)}_{A_{n-1}}(\{\widetilde{t}_i\},\{\widetilde{s}_i\})
 \label{V2I}\end{align} 
where $\varrho$ is the same as in \eqref{varrho-def}. 
The relations \eqref{V1I}, \eqref{W-factors} and \eqref{V2I}
immediately imply that the star-star relation \eqref{star-star} is
equivalent to the Rains transformation formula \eqref{rains-th}
for the elliptic hypergeometric integrals, obtained in \cite{Rains:2010}.

\section*{Acknowledgments}
We are grateful to Hjalmar Rosengren who brought our attention to the
the work of Rains \cite{Rains:2010}. The work was partially supported
by the Australian Research Council.
 
\bibliography{total32,elliptic}
\bibliographystyle{utphys}

\end{document}